# ENHANCED AODV ROUTE DISCOVERY AND ROUTE ESTABLISHMENT FOR QOS PROVISION FOR REAL TIME TRANSMISSION IN MANET

Iftikhar Ahmad[1], Uzma Ashraf[1], Sadia Anum[1], and Hira Tahir[1]

[1]Department of CS & IT Mirpur University of Science and Technology
Mirpur, AJK, Pakistan

## *ABSTRACT*

*MANET is a temporary connection of mobile nodes via wireless links having no centralized base station. We developed a protocol with an enhanced route discovery mechanism that avoids the pre-transmission delay. When a source node wants to communicate with another node, it broadcast RREQ. EAODV give priority to the source node of real time transmission. When RREQ packet send to neighbor node, for real time transmission it accept route request on priority basis and the drop ratio of packets decreased, then throughput increases by receiving more packets at destination and delivery ratio also increased through these QOS improved.*

## 1. INTRODUCTION

Wireless AD-HOC network is distributed network having no centralized infrastructure, each node forward data packets in routing, sometimes other normal nodes may become router or gateway. The first wireless AD-HOC networks were the “packet radio” networks (PRNETs), developed by DARPA in 1970. This network overcome the problem of fixed centralized infrastructure in the environment where single centralized node and fixed nodes via wired links, does not work accurately. This AD-HOC network is very useful in Emergency conditions, ease of deployment, speed of deployment and decreased dependence on infrastructure.

There are three basic types of wireless AD-HOC network: MANET, Wireless mesh network and Wireless sensor network.MANET [1] is a combination of independent mobile nodes that are connected with wireless links (radio waves) to perform peer-to-peer communication. MANET may operate in a standalone fashion, or may be connected to the other larger network like Internet. Applications of MANETs include military use in battle fields, where a centralized command center is not only infeasible but also undesirable; and disaster management scenarios, where on the run, communication between various rescue teams is required in the absence of any existing communication infrastructure. There are three major protocols used for MANET are: Proactive (table driven), Reactive (on-demand) and Hybrid (both proactive and reactive) which are discussed below.

### 1.1 Proactive (table driven)

This type of protocols maintains fresh lists of destinations and their routes by periodically distributing routing tables throughout the network. The main disadvantages of such algorithms include: Respective amount of data for maintenance, slow reaction on restructuring and failures. Examples of proactive are DSDV (destination sequenced distance vector) and OLSR (Optimized Link State Routing).

### 1.2 Reactive (on-demand)

This type of protocols finds a route on demand by flooding the network with Route Request packets. Determine route if and when needed Source initiates route discovery. Examples are DSR (dynamic source routing) and AODV (ad-hoc on demand distance vector).

### 1.3 Hybrid

This category is called as Adaptive; Combination of proactive and reactive. Example ZRP (zone routing protocol)

AODV [2] is efficient for both unicast and multicast routing, it builds routes between nodes only when desired (on-demand) by source nodes, routes are maintain as long as these are needed by source. AODV has four types of messages: RREQ (route request), RREP (route reply), RERR (route error) and hello message.

These messages are used in establishing the route between source and destination. If node wants to send data packet to some destination, it first checks the route to that destination from source, if there is no route is present between source and destination. Then, first by route discovery method route is established between source and destination for data delivery. Source node broadcast RREQ message to its neighbors, node which receive RREQ may send RREP to source if it is destination node or it has route to destination with corresponding sequence number greater than or equal to that contained in the RREQ. Otherwise, it rebroadcasts the RREQ. Nodes keep track of the RREQ's source IP address and broadcast ID. If they receive a RREQ which they have already processed, they discard the RREQ and do not forward it. When source node receive RREP message from destination node route is established between source and destination, then HELLO message generate by source to destination through newly discovered route to check before the data transmission, then source send data through this route to destination. If due to any reason topology change or node die, then link failure occurs and RERR message send to source. After receiving the RERR, if the source node still desires the route, it can reinitiate route discovery.
The rest of this paper is organized as follows. Related work and problem definition is presented in Section II. The solution and design is covered in Section III, followed by implementation results and performance evaluation in Section IV. This paper is concluded in Section V while the references are given towards the end of this paper.

## 2. RELATED WORK

Enhanced Load Balanced AODV Routing Protocol [3] was proposed to overcome the problem for longer period of time a protocol is proposed which selects route on the basis of traffic load on node and reset path as topology changes.

AODVLM [4] which is new load balanced AODV routing protocol, provide a scheme for load distribution among nodes in MANET. AODVLM improves throughput, reduce average end-to-delay and overall network performance.

Energy efficient route discovery method [5] was proposed which reduce the broadcast cost and overall energy of network. In this approach node Discovery process uses regions similar to disk which have variable dimensions and as a result time and energy reduces.

AODV-NC-WLB [6] a modified route request broadcast approach was presented based on node caching. They overcome the drawback of CDS overuse of dominating nodes by a new load balancing scheme in which they measure the protocol fairness using as parameter distribution among nodes of the forwarding load. Work load balance technique is based on the idea by dropping RREQ packet according to load status of each node and load status is set by the value of threshold.

EEMLAR protocol [7] discussed, a node is involved in any transaction, it losses some of its energy whose values depend upon the following factors: nature of packet, size, and distance between nodes. For calculating all these factors, proposed protocol gives an optimized function. After receiving many route request it calls the optimization function to determine the best path to select and send RREP .It also store some inferior paths as backup.

Paper [8] proposed an enhancement in AODV protocol to provide QOS support for real time traffic by using time slot bandwidth reservation in two cases first when both the nodes are in the MANET and second when one node is in MANET or other is in fixed network.

FQMM model [9] (flexible QOS model) consider the characteristics of MANET and combines high quality services of QOS models (intserv and diffserv). FQMM features include: dynamic roles of nodes, hybrid provisioning and adaptive conditioning.

Real Time On-demand Distance Vector in Mobile Ad hoc Networks [10] includes load parameter in the RREQ message that helps in the selection of route with low congestion during route discovery process.

Energy Level and Link State Aware AODV Route Request Forwarding Mechanism Research [11] presented an AODV protocol with enhancements. In route discovery the route selection is made on the basis of node power, load status and link state between.

[12] Combine the shortest route selection criteria of AODV with real network status including link quality, the remaining power capacity and traffic load. This paper contributed by giving status adoptive routing which improves network life and delivery ratio.

[13] Proposed a new QOS routing protocol based on load distribution algorithm intended for a variety of traffic classes to establish the best routing paths. The proposed algorithm calculates the cost metric on the basis of the load on the links. The traffic can be classified as real time traffic and normal traffic. Real time traffic is considered as high priority and normal traffic as low priority. In the absence of real time traffic, the lesser loaded path can be utilized by normal traffic. [14] Presents a comparative analysis of mobile ad-hoc routing protocols over real time video streaming. Their analysis exploits the built-in support for real time multimedia streaming in ad-hoc routing protocols. The performance evaluation has been based on various network level metrics, including average end-to-end delay and packet drop rate.

In the previous techniques there is the problem of RREQ rejection which leads to pre-transmission delay and don't have QOS in real time transmission. When a source node wants to communicate with another node, it sends RREQ to its neighbor nodes, and waits for route reply. If all the neighbor nodes are busy they do not receive RREQ packets and rejected RREQ packets again and again. This leads to the problem of delay in the initial stage of transmission. Resulting into more packet drops and decreased overall throughput. This is called pre -transmission delay.

# 3. SOLUTION AND DESIGN

Our solution to resolve the pre-transmission delay problem is to modify existing AODV route request mechanism in such a way that it will give greater performance then existing AODV protocol. We are enhancing the existing AODV protocol route discovery process and also reduce the pre transmission delay of RREQ in RT transmission. We give priority to the source node of real time transmission. When RREQ packet is broadcasted to neighbor node, for real time transmission it accept route request on priority basis. In basic AODV the route once selected for transmission does not expires until unless the transmission is completed. In our mechanism we have to expire active routes of TCP when UDP transmission starts.As we discussed above that we give priority to real time transmission real time transmission. The drop ratio of packets gradually decreased, then throughput increase by delivering more packets at destination and delivery ratio also increased as a result of which QOS improves. We create scenario in TCL including two flows for TCP (normal packets) and one for UDP (real time packets). We give priority to the source node of UDP connection, when UDP transmission starts there is less drop rate, increased throughput and delivery ratio.

According to EAODV, real time traffic (UDP) gets priority on normal packets. UDP transmission starts at 60 sec in the network, after start of UDP transmission, no TCP transmission starts. UDP has priority to send and receive the packets and in routing table only store information about CBR over UDP transmission.

Table 1: EAODV Mechanism

| Source node | Packet Type | Time to start | Destination node |
|---|---|---|---|
| 1 | TCP | 5.0 sec | 3 |
| 4 | TCP | 35.0 sec | 7 |
| 5 | UDP | 60.0 sec | 8 |

Table 1 shows, there are three flows of transmission of packets. There are two flows of TCP transmission and one flow of UDP transmission. TCP is used to transmit normal packets and UDP is used to transmit real time packets. Source node 1 start transmission at 5.0 sec, its packet type is TCP for normal packets; its destination node is 3. Then, at 35.0 sec source node 4 start transmission of normal packets TCP, its destination node is 7. When at 60.0 sec UDP transmission of real time traffic starts, it has priority with normal packets. After 60.0 sec in routing table, there is no packet of normal traffic. Source node of TCP packet sends their packets but no data received of normal traffic at destination of TCP packets. UDP real time traffic has priority over TCP normal traffic, that's why when UDP transmission starts there is no packet of normal traffic is stored in routing table. As a result, there are more packets of UDP transmission are received at destination, throughput increased, packet loss rate decreased and delivery ratio increased. Hence it is clear that, EAODV is more efficient protocol and also provide QOS qualities in real time transmission.

## 3.1 Enhanced route discovery algorithm

Following is the brief description of our enhanced route discovery algorithm that provides solution to the problem stated earlier in the problem statement.

- **Give priority to source node of special (real time) packets**
- Fill in the essential RREQ fields and
- Broad cost of RREQ Message
- Active route time is set to a constant
- Reception of RREQ.
- **Check If RREQ is special type of packet then expire already store entry in routing table and store special packet on priority basis, accept RREQ.**
- Update the routing entry otherwise do not bother.
- **Forward it to next node.**
- If I am the destination Send RREP message to the updated reverse rout in the table.
- Set the Route expire time to current time + active route timeout
- **Transmission of Real time data packets starts on priority basis.**

# 4. IMPLEMENTATION AND ANALYSIS

We implement EAODV in NS2 [15] and compare its performance with basic AODV. Propagation model is Two Ray Ground and Mobility model is Random Way Point [16] because these models are most widely implemented in simulations of most literature work.

## 4.1 NS-2 Parameter setting

Other parameters of NS2 are configured for our network scenario is shown in the Table 2. NS2 is discrete network simulator that is used for implementation for performance comparison MANET protocols

Table 2. NS-2 Parameter Setting

| | |
|---|---|
| Routing protocol | AODV |
| Link layer type | LL |
| MAC type | Mac/802_11 |
| Interface Queue | Queue/Drop Tail/PriQueue |
| Maximum packets in Ifq | 20 |
| Antenna model | Antenna/Omni Antenna |
| Radio propagation model | Propagation/TwoRay Ground |
| Network Interface | Phy/WirelessPhy |
| Channel type | Channel/Wireless Channel |
| Number of mobile nodes | 10 |
| Simulation time | 150 seconds |

## 4.2 Performance Metric

We will compare our enhanced protocol with basic AODV. We evaluates performance mainly according to following metric.

### 4.2.1 Average throughput

Throughput is the total number of packets received successfully in the given time.

### 4.2.2 Packet delivery ratio

It is the ratio of the number of packets received successfully to the total number of packets transmitted.

### 4.2.3 Packet loss

It is the packet loss rate of the transmission**.**

## 4.3 Simulation results

The Basic AODV is first implemented and run on given MANET scenario in NS-2 and performance parameters are calculated. Results are generated in graphical form produced by using trace files that is generated during simulation. Then our proposed protocol for real time traffic is implemented and results are produced in graphical form.Results are compared by taking pause time on X-axis and performance metrics throughput and delay on Y-axis.

## 4.4 Comparison of performance

We simulated the results in NS2 by taking parameters (pause time) on x-axis and performance metric (delivery ratio, packet loss rate and throughput) on Y-axis. The pause time is the time interval during which different parameters are calculated .In the figure below at equal intervals the throughput is calculated as shown in figure 1.

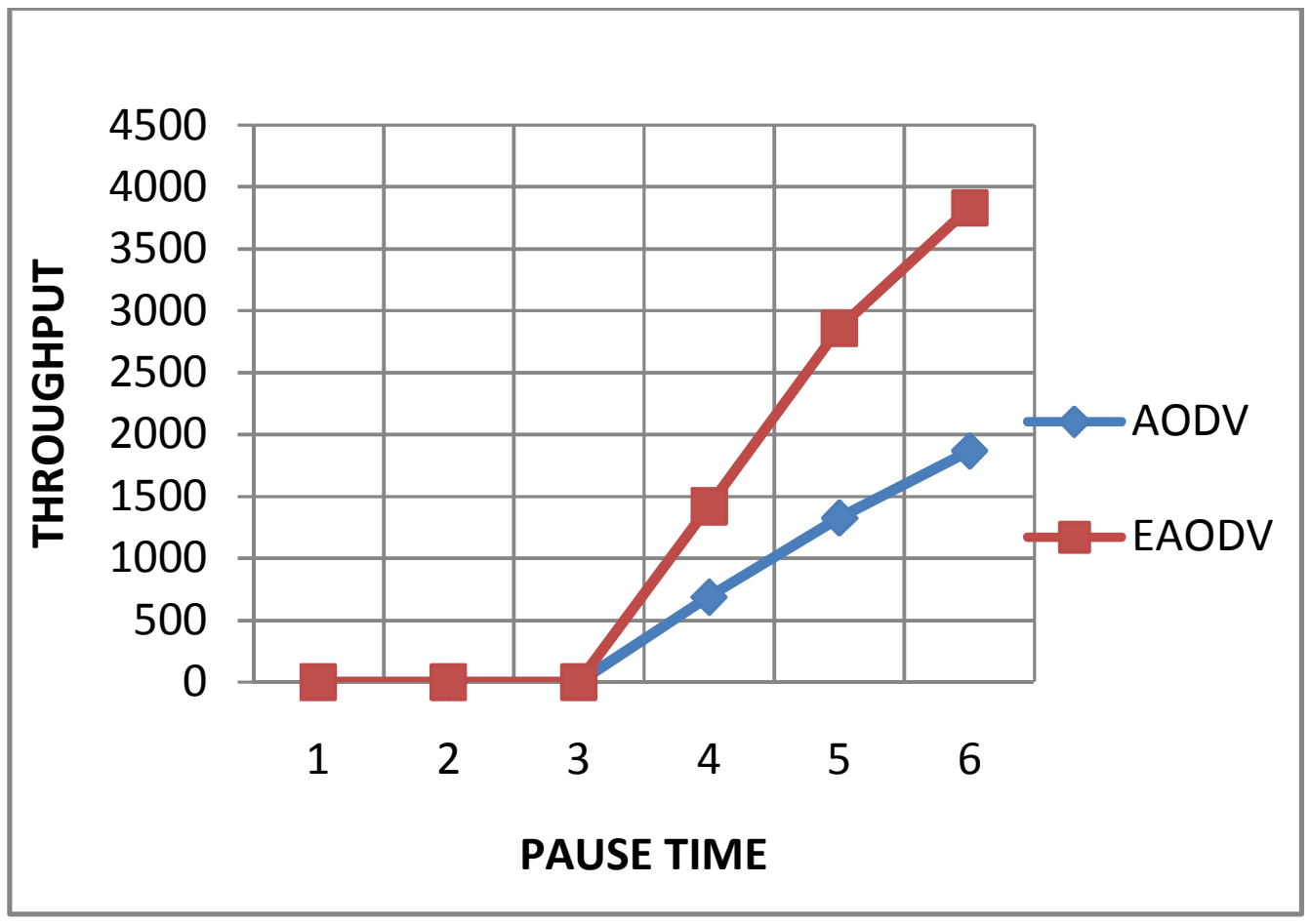

Figure 1. Comparison of throughput of AODV and EAODV

In figure 1 the throughput comparison between AODV and EAODV is given and it is clear that the throughput of EAODV is greater than basic AODV as indicated by red line in the graph. At point 1, 2 and 3 throughput of CBR packets is zero because UDP transmission is not started yet. At point 3 UDP transmission start and a large increase occur in throughput of CBR packets during UDP transmission. After point 3 graphs shows a continuous increase in throughput rate of CBR packets as compare to basic AODV.

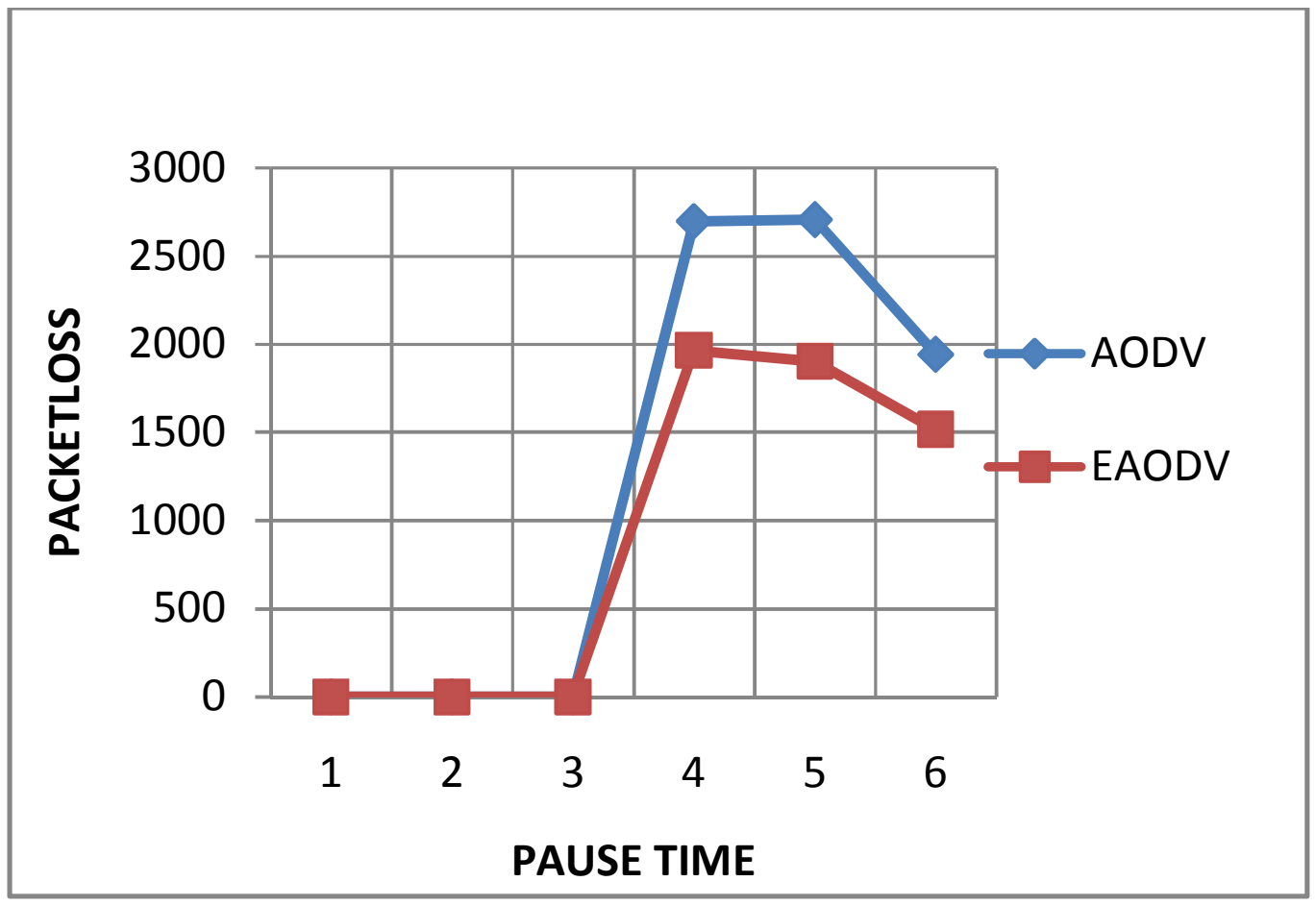

Figure 2. Comparison of packet loss of AODV and EAODV

The packet loss rate of basic AODV and EAODV is compared in figure 2. The packet loss rate of both protocols is calculated at fixed interval of time. First we calculated the number of packets during fixed intervals and packet loss then packet loss rate is calculated. At point 1, 2 and 3 packet loss is zero because UDP transmission is not start at that time. At point 3 UDP transmission start and a large decrease occur in packet loss of CBR packets during UDP transmission as compare to basic AODV. The packet loss rate for the EAODV is smaller than basic AODV.

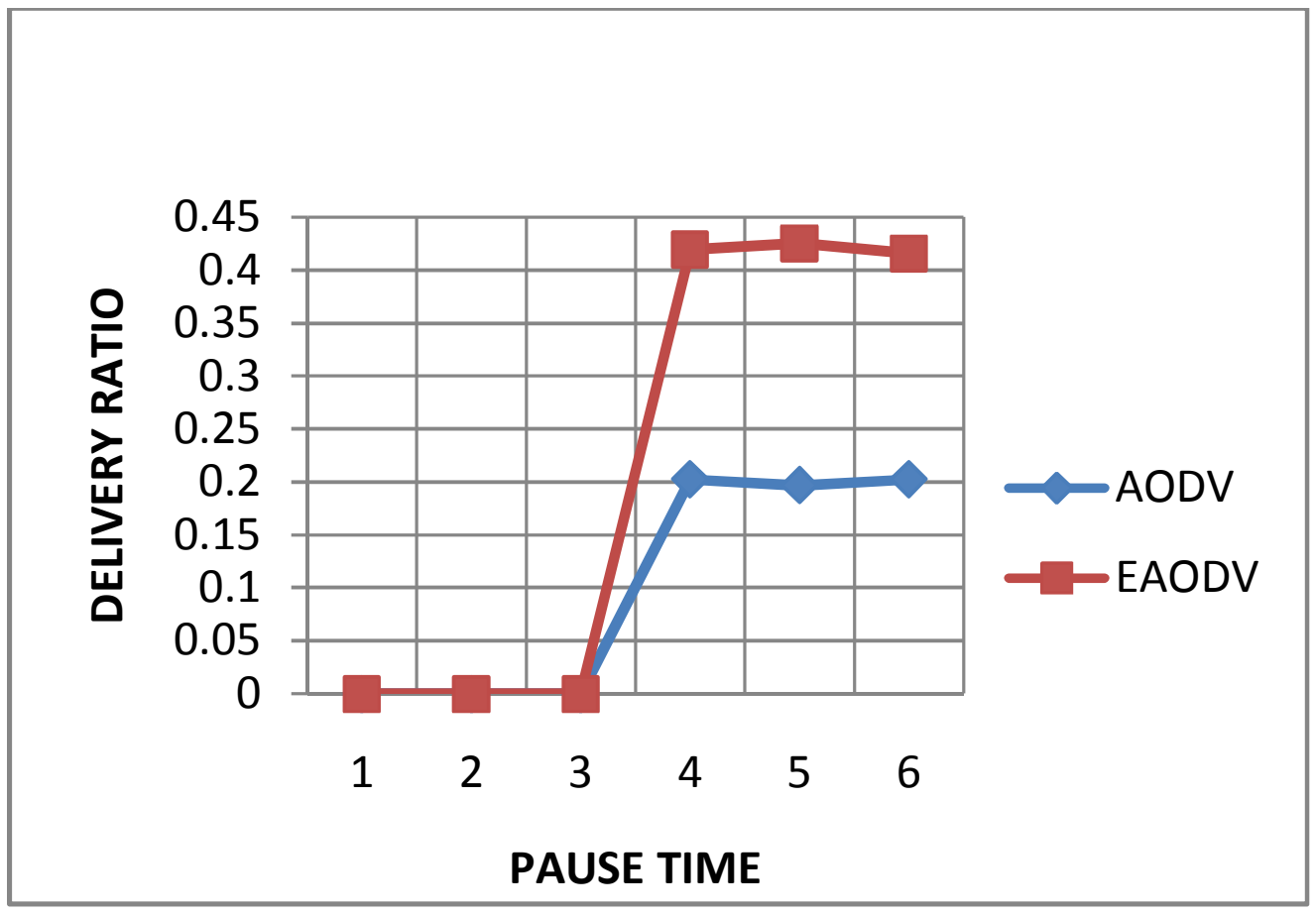

Figure 3. Comparison of delivery ratio of AODV and EAODV

Figure 3 shows the delivery ratio of both protocols. Red line is indicating the delivery ratio of EAODV and blue line indicates the basic AODV. At point 1, 2 and 3 delivery ratio is zero because UDP transmission is not start. At point 3 UDP transmission start and a large increase occur in delivery ratio of CBR packets during UDP transmission. Graphs shows continuous increase in delivery ratio because during UDP transmission no TCP packets are send or received, due to priority only CBR packets are transmitted. Hence delivery ratio of CBR packets is increased.

## 5. CONCLUSION

We developed a protocol with an enhanced route discovery mechanism that avoids the RREQ rejection and in results reduces the pre transmission delay. EAODV give priority to the source node of RT transmission. When RREQ packet sends to neighbor node, for RT transmission it accept route request on priority basis and starts the RT transmission. EAODV expire active routes of TCP when UDP transmission starts.

The drop ratio of packets decreased, then throughput increase by receiving more packets at destination and delivery ratio also increased. Our technique focuses to maintain real time transmissions. With these enhancements our protocol increases the overall throughput of network and packet delivery ratio for real time transmission as compare to the basic AODV routing protocol. The simulation result proves the improvement in the performance of the protocol.